\documentstyle[aps]{revtex}

\newcommand{\be}{\begin{equation}}
\newcommand{\ee}{\end{equation}}
\newcommand{\bea}{\begin{eqnarray}}
\newcommand{\eea}{\end{eqnarray}}
\newcommand{\bml}{\begin{mathletters}}
\newcommand{\eml}{\end{mathletters}}
\newcommand{\pa}{\partial}

\newcommand{\dxy}{\delta(\vec{x}-\vec{y})}

\newcommand{\vx}{\vec{x}}
\newcommand{\vy}{\vec{y}}

\newcommand{\vk}{\vec{k}}

\newcommand{\e}{\epsilon}
\newcommand{\ve}{\varepsilon}

\begin{document}
\draft
\title{DUALITY SYMMETRY IN THE SCHWARZ-SEN MODEL\footnote{Talk delivered
at ``TRENDS IN THEORETICAL PHYSICS, SECOND CERN-SANTIAGO DE COMPOSTELA-LA PLATA
MEETING'', Buenos Aires, Argentina (1998).}}
\author{H. O. Girotti}
\address{Instituto de F\'{\i}sica,
Universidade Federal do Rio Grande do Sul \\ Caixa Postal 15051, 91501-970  -
Porto Alegre, RS, Brazil.}

\maketitle

\begin{abstract}
The duality symmetric but non manifestly covariant action proposed by Schwarz 
and Sen is canonically quantized in the Coulomb gauge. The resulting theory 
turns out to be, nevertheless, relativistically invariant. It is shown, 
afterwards, that the Schwarz-Sen model naturally emerges when duality is
implemented as a local symmetry of sourceless electrodynamics. This implies in
the equivalence of these theories at the quantum level. 
\end{abstract}

\newpage

\section{\bf Introduction}

The equations of motion of the four-dimensional low energy effective field 
theory for the bosonic sector of the heterotic string, are invariant under 
SL(2,R) duality transformations of the massless fields involved. This, so 
called S-duality symmetry, is a symmetry of the equations of motion but 
not of the corresponding action. On the other hand, the low energy effective 
field theory action retains the target space duality symmetry (T-duality)
of string theory. It would be desirable to achieve S-duality at the level of 
the action in the hope of attaining results similar to those given by the 
T-duality symmetry. However, it is difficult to conciliate duality symmetry and
manifest Lorentz covariance.

The difficulty of writing a non trivial action involving only a 
finite-component self-dual covariant form is well known. In fact, to construct
a duality symmetric action being manifestly covariant one 
must introduce auxiliary fields. The first attempts in this direction involved
an infinite number of auxiliary fields\cite{McClain}. 

Duality symmetric actions being manifestly covariant and containing a finite 
number of auxiliary fields have also been constructed but they are not
polynomials. These actions were found by Pasti, Sorokin and Tonin 
(PST)\cite{PST1,KP}. For the case of electrodynamics, the
PST action covariantizes the action found earlier by Deser and Teitelboim
\cite{SD1} and rediscovered by Schwarz and Sen\cite{SSen}, which is duality 
symmetric but not manifestly covariant. The introduction of
sources into the manifestly covariant and non-manifestly covariant versions of
these duality symmetric actions has also been achieved\cite{Nathan2}.

The present work is essentially dedicated to establish the quantum mechanical
equivalence of the Schwarz-Sen and Maxwell theories in the case free of
sources. We start by quantizing the Schwarz-Sen
model in the Coulomb gauge. The resulting quantum theory will be shown to 
be, nevertheless, relativistically invariant. This is our Section II. In 
Section III we begin by recalling that the Maxwell action, when formulated in 
the Coulomb gauge, remains invariant under a set of non-local duality
transformations derived by Deser and Teitelboim \cite{SD1}. Additional 
fields are, afterwards, brought into the theory in order to make these 
transformations local. Correspondingly, a new expression for the generating 
functional of Green functions is derived. In Section IV we demonstrate that 
the Green functions generating functionals for the Maxwell and Schwarz-Sen 
theories are rigorously identical. Section V contains the conclusions.    
 
\section{\bf Quantization of the Schwarz-Sen Model}

The duality symmetric action proposed by Schwarz-Sen\cite{SSen} involves two 
gauge potentials $ A^{\mu,a}(0\le{\mu}\le{3},\,\, 1\le{a}\le{2})$ and 
reads\footnote{This section is mainly based on Refs.\cite{Gi1,Gi2}. Our 
space-time metric is $g_{00}=-g_{11}=-g_{22}=1$ while $\epsilon_{ab}$ 
designates a generic element of the two-dimensional antisymmetric unit matrix 
($\epsilon_{12}=+1$). } 

\be
\label{10}
S = - \frac{1}{2} \int d^4x \left( B^{a,i} \epsilon_{ab} E^{b,i} \,+\,
B^{a,i}B^{a,i} \right)\,\,\,,
\ee

\noindent
where

\bml
\label{101}
\bea
&& E^{a,i} = - F^{a,0i} = - \left( \partial^{0}A^{a,i} -
\partial^{i}A^{a,0}\right)\,\,\,,\label{mlett:a101}\\
&& B^{a,i} = -\frac{1}{2} \epsilon^{ijk}
F^{a}_{jk} = -\epsilon^{ijk}\partial_{j}A^{a}_{k}\,\,\,\label{mlett:b101},
\eea
\eml

\noindent
and $1\le{i,j,k}\le{3}$. $S$ is separately invariant under the local gauge 
transformations

\bml
\label{102}
\bea
&& A^{a,0} \rightarrow A^{a,0} + \Psi^{a}\,\,\,,\label{mlett:a102}\\
&& A^{a,i} \rightarrow A^{a,i} - \partial^{i}\Lambda^{a}\,\,\,,
\label{mlett:b102} 
\eea
\eml

\noindent
and under the global $SO(2)$ rotations

\be
\label{103}
A^{\mu a} \rightarrow A^{\prime\mu a} =  
A^{\mu a}\cos \theta + \epsilon^{a b} A^{\mu b} \sin \theta\,\,\,.
\ee

\noindent
Of course, (\ref{103}) reduces to the usual discrete duality transformation for
$\theta = \pi/2$. However, the Lagrangian density in (\ref{10}),

\be
\label{11}
{\cal{L}} = \frac{1}{2} \e^{jki}(\pa_{j}A^{a}_{k})\e_{ab}(\pa_{0}A^{b}_{i})-
\frac{1}{2} \e^{jki}(\pa_{j}A^{a}_{k})\e_{ab}(\pa_{i}A^{b}_{0})
-\frac{1}{4} F^{a,jk} F^a_{jk}\,\,\,,
\ee

\noindent
is not a Lorentz scalar. The use of the equations of motion deriving from 
(\ref{11}),

\be
\label{12}
\epsilon^{ijk} \epsilon_{ab} \pa_{0} \pa_{j} A^{b}_{k} + \pa_{j}
\left(\pa^{j} A^{a,i} - \pa^{i} A^{a,j}\right) = 0\,\,\,,
\ee

\noindent
allows for the elimination from $S$ of one of the gauge fields, the action for
the remaining one being the conventional Maxwell action.

Within the Hamiltonian framework, the Schwarz-Sen model is characterized by the
canonical Hamiltonian ($H_c$)

\be
\label{13}
H_c = \int d^{3}x \left[ \frac{1}{2} \e^{jki}(\pa_{j}A^{a}_{k})\e_{ab}
(\pa_{i}A^{b}_{0}) + \frac{1}{4} F^{a,jk}F^a_{jk} \right]\,\,\,.
\ee

\noindent
Furthermore, the system possesses the primary constraints

\bml
\label{14}
\bea
\Omega^a_0 &\equiv& \pi^a_0 \approx 0\,\,\,, \label{mlett:a14}\\
\Omega^a_i &\equiv& \pi^a_i + \frac{1}{2} \e_{ab}\,\e_{ijk}\,\pa^j A^{b,k}
\approx 0\,\,\, , \label{mlett:b14}
\eea
\eml

\noindent
where we have designated by $\pi^a_{\mu}$ the momentum canonically conjugate to
$A^{a,\mu}$. Then, the total Hamiltonian ($H^{\prime}$) is given by
$H^{\prime} = H_c + \int d^3x \left( u^{a,0} \Omega^a_0 + 
u^{a,i} \Omega^a_i \right)$,
where the $u$'s are Lagrange multipliers. Persistence in time of $\Omega^a_0$
produces neither secondary constraints nor determines the Lagrange
multipliers. On the other hand, persistence in time of the primary constraints
$\{\Omega^a_i\}$ does not lead to the existence of secondary constraints but 
determines partially the Lagrange multipliers $\{u^a_i\}$. Indeed, since 
the Poisson bracket

\be
\label{15}
[\Omega^a_i(\vec{x})\, , \, \Omega^b_j(\vec{y})]_{P} = - \e_{ab}\,\e_{ijk} 
\, \pa^j_x \dxy
\ee

\noindent
does not vanish, ${\dot{\Omega}}^a_i = [\Omega^a_i , H^{\prime}]_{P} 
\approx 0 $ yields $u^{a,i} = \e_{ab}(B^{b,i} -
\pa^i\phi^{b})$, where $\phi^{a}$ is an arbitrary scalar. Thus, 

\be
\label{16}
\Omega^{a}(\vec{x}) = \pa^i \Omega^a_i(\vec{x}) \approx 0
\ee

\noindent
and $\Omega^a_0 \approx 0 $ are the first-class constraints in the 
theory. 

To isolate the second-class constraints from (\ref{mlett:b14}), we split 
$\Omega^a_i$ into longitudinal ($L$) and
transversal ($T$) components, namely, $\Omega^a_i = \Omega^a_{Li} 
+ \Omega^a_{Ti}$, where $\Omega^a_{Li} = -\frac{\pa_i \pa^j}{\nabla^2}
\Omega^a_j$, $\Omega^a_{Ti} = \left(g_i^j 
+ \frac{\pa_i \pa^j}{\nabla^2}\right)\Omega^a_j$ and 
$\nabla^2 \equiv -\pa_j \pa^j $. The first-class constraint (\ref{16}) 
only involves the longitudinal components $\Omega^a_{Li}$ and states that these
components vanish individually. Then, the second-class constraints are
 
\be
\label{17}
\Omega^a_{Ti} = \pi^a_{Ti} + \frac{1}{2} \e_{ab}\,\e_{ijk}\,\pa^j A^{b,k}_T
\approx 0\,\,\,.
\ee

The determination of the constraint structure is over. It only 
remains to be mentioned that the gauge potential $A^{a,\mu}$, when acted 
upon by the generator of infinitesimal gauge transformations, 
$G = \int d^3x \left(\Psi^a \Omega^a_0 + \Lambda^a \Omega^a\right)$,
undergoes the change $A^{a,\mu}\rightarrow A^{a,\mu}+\delta A^{a,\mu}$
with $\delta A^{a,0} = [A^{a,0},G]_{P} = \Psi^a$ and
$\delta A^{a,i} = [A^{a,i},G]_{P} = - \pa^i \Lambda^a$, in agreement with
Eqs.(\ref{102}).

We shall next quantize the theory by means of the Dirac bracket quantization
procedure\cite{Di,Fr,Su,Gi}. To this end, we start by fixing the gauge through
the subsidiary conditions

\bml
\label{18}
\bea
&&\chi^{a,0} \equiv A^{a,0} \approx 0\,\,\,,\label{mlett:a18}\\
&&\chi^a \equiv \pa_iA^{a,i} \approx 0 \,\,\,.\label{mlett:b18}
\eea
\eml

\noindent
The fact that the Coulomb condition and $A^{a,0} \approx 0$ are, when acting
together, accessible gauge conditions is a peculiarity of the model under
analysis. We now recall that, according to the quantization procedure being 
used, the equal-time commutation algebra is to be abstracted from the 
corresponding Dirac bracket algebra, the constraint and gauge conditions
thereby translating into strong operator relations. After some calculations one
finds that

\bml
\label{19}
\bea
&& \left [ A^{a,i}_T(\vx ) \,,\, A^{b,j}_T(\vy ) \right ]\,
=\,-i \e_{ab}\,\e^{ijk}
\frac{\pa^x_k}{\nabla^2}\dxy\,\,\,, \label{mlett:a19}\\
&& \left [ A^{a,i}_T(\vx ) \,,\, \pi^b_{Tj}(\vy ) \right ]\,=\, \frac{i}{2}
\delta_{ab}\left( g^i_j + \frac{\pa^i_x \pa_j^x}{\nabla^2} \right)\dxy \,\,\,,
\label{mlett:b19}\\ 
&& \left [ \pi^{a}_{Ti}(\vx) \,,\, \pi^b_{Tj}(\vy) \right ]\,=\, \frac{i}{4}
\e_{ab}\, \e_{ijk} \pa^k_x \dxy \,\,\,, \label{mlett:c19} 
\eea
\eml

\noindent
while the Hamiltonian operator reads

\be
\label{20}
H = \frac{1}{4} \int d^3x  F^{a,jk}F^a_{jk}\,=\,-\frac{1}{2}\int d^3x 
B^{a,j}B^a_j \,\,\,.
\ee

\noindent
One may wonder on whether the right hand side of (\ref{20}) is afflicted by 
ordering ambiguities. This not so, since

\be
\label{21}
\left[B^{a,i}(\vx)\,,\,B^{b,j}(\vy)\right]\,=\,i\,\e_{ab}\,\e^{ijk}\,
\pa_k^x\dxy \,\,\,.
\ee

The main object of interest is the field commutator at different space-time 
points. To find it, we must first solve the Heisenberg equations of motion 
deriving from (\ref{19}) and (\ref{20}), namely,

\bml
\label{22}
\bea
&& {\cal {D}}^{(-)ab}_{ik} A^{b,k}_T \,=\,0 \,\,\,,\label{mlett:a22}\\
&& \pa_0 \pi^a_{T i} \,=\,\frac{1}{2} \pa^j F^a_{ji}\,\,\,,\label{mlett:b22} 
\eea
\eml

\noindent
where

\be
\label{23}
{\cal {D}}^{(\pm) ab}_{ik}\equiv g_{ik} \delta_{ab}\pa_0\,
\pm \,\e_{ab}\e_{ijk}\pa^j \,\,\,.
\ee

\noindent
Notice that, in the Coulomb gauge, the Lagrange equation of motion (\ref{12})
can be casted as

\be
\label{24}
\e^{jli} \pa_l \, {\cal {D}}^{(-) ab}_{ik} A^{b,k}_T \,=\,0 \, \Longrightarrow 
 {\cal {D}}^{(-) ab}_{ik} A^{b,k}_T \,=\,\pa_i \xi^a \,\,\,.
\ee

\noindent
Since $\pa^i {\cal {D}}^{(-) ab}_{ik} A^{b,k}_T \,=\,0$, 
the function $\xi^a$ must
verify $\nabla^2 \xi^a = 0$ but is otherwise arbitrary. Thus, the Lagrangian 
and the Hamiltonian formulations lead to equivalent equations of motions only 
after the introduction of a regularity requirement at spatial infinity.  This 
situation resembles that encountered in connection with the theory of the 
two-dimensional ($x^0, x^1, x^{\pm} = 1/\sqrt{2}(x^0 \pm x^1)$) self-dual field
($\Phi$) proposed by Floreanini and Jackiw\cite{Ja1,Gi3}, 
where the equations of motion in the Lagrangian and Hamiltonian formulations 
turn out to be, respectively, $\pa_1 \pa_{-}\Phi = 0$ and $\pa_{-}\Phi = 0$. 
We also recall that in order to solve $\pa_{-}\Phi = 0$ one starts by 
realizing that
$\pa_{-}\Phi = 0 \Longrightarrow \pa _{+} \pa_{-}\Phi = 0 \Longrightarrow 
\Box \Phi = 0$. The solutions of $\pa_{-}\Phi = 0$ are then contained in the
field of solutions of $\Box \Phi = 0$. We shall follow here a
similar approach, since

\be
\label{25}
{\cal {D}}^{(-) ab}_{ik} A^{b,k}_T \,=\,0 \Longrightarrow 
{\cal {D}}^{(+) ca,li} {\cal {D}}^{(-) ab}_{ik} A^{b,k}_T \,=\,0 
\Longrightarrow \Box A^{c,l}_T = 0\,\,\,. 
\ee        

\noindent
The solving of $\Box A^{a,i}_T = 0$ leads to

\be
\label{26}
A^{a,i}_T(x)\,=\,\int d^3y D(x - y){\buildrel{\scriptstyle{\leftrightarrow}}
\over{\textstyle{\partial}}}^0_y A^{a,i}_T(y)\,\,\,,
\ee

\noindent
where $D(x -y)$ is the zero-mass Pauli-Jordan delta function and
$(A{\buildrel{\scriptstyle{\leftrightarrow}}  \over{\textstyle{\partial}}
}^k B) \equiv A \partial^k B - B \partial^k A $. The combined use of this
last equation and (\ref{19}) allowed us to find the following explicit form for
the field commutator at different space-time points 

\bea
\label{27}
&&\bigl[A^{a,i}_T(x)\,,\,A^{b,j}_T(y) \bigr] \nonumber\\
&&=\,i\,\left[\delta_{ab} 
\left( g^{ij} + \frac{\pa^i_x \pa^j_x}{\nabla^2_x} \right)
\,-\,\e_{ab} \e^{ijk} \frac{\pa^x_k \pa^x_0}{\nabla^2_x} \right]\,
D(x - y)\,\,\,. 
\eea

\noindent
One can verify, by applying ${\cal {D}}^{(-) ca}_{ki}(x)$ to both sides of 
(\ref{27}), that the field configurations entering the just mentioned 
commutator are in fact solutions of (\ref{mlett:a22}).

Now, the function $D(x - y)$ can be given as the sum of a positive plus a
negative frequency part and we, therefore, can write

\be
\label{28}
A^{a,i}_T(x) \,=\,A^{a,i (+)}_T(x)\,+\,A^{a,i (-)}_T(x)\,\,\,,
\ee
  
\noindent
where

\bea
\label{29}
&&A^{a,i (\pm)}_T(x) \nonumber\\ 
&&=\,\frac{1}{(2\pi)^{3/2}}\int
\frac{d^3k}{\sqrt{2|\vec{k}|}} \exp [\pm i(|\vec{k}|x^0 - \vec{k}\cdot \vx)]
\sum_{\lambda = 1}^{2}\ve^{a,i}_{\lambda}(\vec{k})a^{(\pm)}_{\lambda}(\vk)
\,\,\, 
\eea

\noindent
and $\ve^{a,i}_{\lambda}(\vec{k}), \lambda = 1,2,$ are unit norm 
polarization vectors. By going back with (\ref{29}) into (\ref{27}) one 
obtains

\bea
\label{30}
&&\sum_{\lambda,\lambda^{\prime} = 1}^{2} \ve^{a,i}_{\lambda}(\vec{k})
\ve^{b,j}_{\lambda^{\prime}}(\vec{k^{\prime}})\left[a^{(-)}_{\lambda}(\vk)\,,\,
a^{(+)}_{\lambda^{\prime}}(\vec{k^{\prime}})\right]\nonumber\\ 
&&= \left[-\delta_{ab}
\left(g^{ij} + \frac{k^i k^j}{|\vk|}\right)\,+\,\e_{ab}\e^{ijl}\frac{k_l}
{|\vk|}\right]\delta(\vk - \vec{k^{\prime}})\,\,\,,
\eea

\noindent
while all others commutators vanish. The polarization vectors are to be found
by replacing (\ref{29}) into the gauge condition (\ref{mlett:b18}) and the
equation of motion (\ref{mlett:b22}). One arrives to

\be
\label{31}
\sum_{\lambda = 1}^{2} {\vec{\ve}}^{\,\, a}_{\lambda}(\vk)\, \times\, 
{\vec{\ve}}^{\,\, b}_{\lambda}(\vk) \,
= \, -\, 2\, \e_{ab} \,\frac{\vk}{|\vk|}\,\,\,.
\ee

\noindent
On the other hand, the Coulomb
gauge polarization vectors span, by construction, the space orthogonal to 
$\vk$, i.e.,

\be
\label{32}
\sum_{\lambda = 1}^{2} \ve^{\,\, a,i}_{\lambda}(\vk) \,
\ve^{\,\, a,j}_{\lambda}(\vk) \,
= \,-\left( g^{ij}\,+\,\frac{k^i k^j}{|\vk|^2}\right) \,\,\,.
\ee

\noindent
By using (\ref{31}) and (\ref{32}) we solve at once for the commutator 
in (\ref{30}),

\be
\label{33}
\left[a^{(-)}_{\lambda}(\vk)\,,\,
a^{(+)}_{\lambda^{\prime}}(\vec{k^{\prime}})\right]\,=\,\delta_{\lambda
\lambda^{\prime}}\, \delta(\vk\,-\,\vec{k^{\prime}})\,\,\,.
\ee

\noindent
Thus the space of states is, as expected, a Fock space with positive definite 
metric. 

Hence, the quantized Schwarz-Sen model is a physically sensible quantum field 
theory. Our next task is to demonstrate that this theory is also 
relativistically invariant. We are therefore looking for a set of 
composite operators $\{\Theta_{\mu \nu}\}$ which may serve as Poincar\'e 
densities. We shall build them by following the rules that are used to
construct the symmetric (Belinfante) energy-momentum tensor in a manifestly 
Lorentz invariant theory. In this way we find

\bml
\label{34}
\bea
\Theta_{00}\,&=&\,-
\frac{1}{2}\,B^{a,i}\,B^{a}_{i}\,\,\,,\label{mlett:a34}\\
\Theta_{0i}\,&=&\,\Theta_{i0}\,=\,-
\frac{1}{2}\,\e_{ijk}\,\e_{ab} B^{a,j}\,B^{b,k}\,\,\,,\label{mlett:b34}\\
\Theta_{ij}\,&=&\,\Theta_{ji}\,=\,-  B^{a}_{i}\,B^{a}_{j}\,+\,g_{ij}\,
 B^{a,l}\,B^{a}_{l}\,\,\,.\label{mlett:c34} 
\eea
\eml

\noindent
Thus, $\Theta$ is symmetric and free of ordering ambiguities but we can not yet
decide on whether or not it is a tensor. As for the equal-time commutator 
algebra obeyed by the components of $\Theta $, it is fully determined
by the commutator (\ref{21}). In particular, one can corroborate that 

\bml
\label{35}
\bea
&&\left[\,\Theta^{00}(x^0,\vec{x})\,,\,\Theta^{00}(x^0, \vec{y})\,\right]
\nonumber\\
&&= -\,i\,\left\{\,\Theta^{0k}(x^0,\vec{x})\,
+\,\Theta^{0k}(x^0, \vec{y})\,\right\} \,\partial^x_k\delta(\vec{x}-\vec{y})\,
\,\,,\label{mlett:a35}\\
&&\left[\,\Theta^{00}(x^0,\vec{x})\,,\,\Theta^{0k}(x^0, \vec{y})\,\right]
\nonumber\\
&&= \,-\,i\,\left\{\,\Theta^{kj}(x^0,\vec{x})\,-\,g^{kj}\,
\Theta^{00}(x^0, \vec{y})
\,\right\}\, \partial^x_j \,{\delta(\vec{x}-\vec{y})}
\,\,\,,\label{mlett:b35}\\
&&\left[\,\Theta^{0k}(x^0,\vec{x})\,,\,\Theta^{0j}(x^0, \vec{y})\,\right]
\nonumber\\
&&= i\,\left\{\,\Theta^{0k}(x^0, \vec{y})\,\partial^j_x\,
+\,\Theta^{0j}(x^0,\vec{x})\,\partial^k_x\,\right\}\,{\delta(\vec{x}-\vec{y})}
\,\,\,. \label{mlett:c35}
\eea
\eml

\noindent
As known\cite{BD}, positivity requires that a singular Schwinger term,
proportional to $(\partial^3)\delta (\vec{x} - \vec{y})$, must also be present
in the right hand side of Eq.(\ref{mlett:b35}). The definition of 
$\Theta^{\mu\nu}$ can, in fact, be altered as to yield such term. But, 
since Schwinger terms do not contribute to the algebra of integrated charges, 
we have omitted them in Eqs.(\ref{35}). From Eqs.(\ref{35}) then follows 
that the charges

\bml
\label{36}
\bea
&& P^{\mu}\,\equiv\,\int d^3x\,\Theta^{0 \mu}\,\,\,, \label{mlett:a36}\\
&& J^{\mu \nu}\,\equiv\,\int d^3x \left(\Theta^{0 \mu} x^{\nu}\,-\,
\Theta^{0 \nu} x^{\mu}\right)\,\,\,,\label{mlett:b36}
\eea
\eml

\noindent
obey the Poincar\'e algebra, i.e.,

\bea
\label{37}
&&\left[P^{\mu} , P^{\nu}\right] = 0\,\,\,,\\
&&\left[J^{\mu \nu} , P^{\sigma}\right] = i \left(g^{\mu \sigma} 
P^{\nu}\,-\,g^{\nu \sigma} P^{\mu}\right)\,\,\,,\\
&&\left[J^{\mu \nu} , J^{\rho \sigma}\right] = i
\left( g^{\mu \rho}J^{\nu \sigma} + g^{\nu \sigma}J^{\mu \rho} -
g^{\mu \sigma}J^{\nu \rho} - g^{\nu \rho}J^{\mu \sigma}\right)\,\,\,.
\eea

\noindent
It takes just a few more steps to demonstrate that $\Theta $ is a
tensor.  Indeed, the additional equal-time commutators 
$\left[\,\Theta^{ij}(x^0,\vec{x})\,,\,\Theta^{00}(x^0, \vec{y})\,\right]$ and
$\left[\,\Theta^{ij}(x^0,\vec{x})\,,\,\Theta^{0k}(x^0, \vec{y})\,\right]$ can
also be readily evaluated by using (\ref{34}) and (\ref{21}). These results 
and (\ref{35}) can be collected into

\bml
\label{38}
\bea
&& \left[P^{\mu}\,,\,\Theta^{\alpha \beta}\right] =
-i\,\pa^{\mu}\Theta^{\alpha \beta}\,\,\,,\label{mlett:a38}\\
&&\left[J^{\mu \nu}\,,\,\Theta^{\alpha \beta}\right] =
-i\left(x^{\nu}\pa^{\mu} - x^{\mu}\pa^{\nu}\right)\Theta^{\alpha \beta}
\nonumber\\
&&-i\left(\Theta^{\mu \alpha}g^{\nu \beta} +
\Theta^{\mu \beta}g^{\nu \alpha} - \Theta^{\nu \alpha}g^{\mu \beta} -
\Theta^{\nu \beta}g^{\mu \alpha}\right)\,\,\,,\label{mlett:b38}
\eea
\eml

\noindent
which are, respectively, the translation and rotation transformation laws to be
obeyed by a second-rank tensor. The purported proof of relativistic
invariance of the quantized Schwarz-Sen theory is now complete. 

What remains to be done is to demonstrate that the Coulomb gauge formulation of
the quantized Schwarz-Sen theory is in fact covariant. Since translations and 
ordinary rotations do not destroy the Coulomb gauge condition we concentrate 
on Lorentz boosts. By using (\ref{36}), (\ref{34}), (\ref{mlett:b101}) and 
(\ref{mlett:a19}) one arrives to

\be
\label{39}
- i \left[J^{0k}\,,\,A^{a,i}_{T}\right]\,=\,(x^0\,\pa^{k}
\,-\,x^{k}\,\pa^{0})A^{a,i}_{T}\,-\,
\e_{ab}\e^{klj} \frac{\pa^{i}\pa_{l}}{\nabla^2} A^{b}_{T j}\,\,\,.
\ee

\noindent
The term proportional to $\e_{ab}$ signalizes that gauge potentials 
corresponding to different values
of $a$ get mixed by Lorentz boosts. This does not occur for ordinary 
rotations. Furthermore, the mixing term in (\ref{39}) describes an operator 
gauge transformation, which, as one easily verifies, makes 
this commutator compatible with the transversality condition $\pa_i A^{a,i}_T 
= 0$. Hence, under Lorentz boosts, the field $A^{a,i}_{T}$ undergoes, besides 
the usual vector transformation, an operator gauge transformation 
which restores the Coulomb gauge in the new Lorentz frame.
  
As for the Nother's charge associated with the $SO(2)$ symmetry (\ref{103}), it
is straightforward to verify that it can be written as 

\begin{equation}
\label{41}
Q=-\frac{1}{2}\int d^3 x \epsilon^{jik} (\partial_j A^{a}_{Ti}) A_{T k}^{a} = 
\frac{1}{2}\int d^3x B^{a k} A^{a}_{T k}\,\,\,.
\end{equation}

\noindent
Observe that $Q$ is a SO(2) invariant Chern-Simons term. Thus, up to
surface terms, it is gauge invariant. It is also metric independent
and so its algebraic form also holds for curved spaces. The use of
(\ref{mlett:a19}) enables one to verify that $Q$ indeed generates the 
infinitesimal $SO(2)$ rotations

\begin{equation}
\label{42}
[Q, A^{b}_{Tj}(y)]= - i \epsilon^{ba} A^{a}_{Tj}(y)\,\,\,.
\end{equation}

\noindent
Furthermore, in terms of the creation and anhilation operators of (\ref{29})
the operator $Q$ is found to read

\begin{equation}
\label{43}
Q= i\int d^3k (a^{\dag}_1 a_2 - a^{\dag}_2 a_1)
\end{equation}

\noindent
and becomes diagonal, 

\begin{equation}
\label{44}
Q=\int d^3 k (a^{\dag}_{L} a_L - a^{\dag}_{R} a_R)\,\,\,,
\end{equation}

\noindent
in the base of circularly polarized operators, defined by

\bml
\label{45}
\bea
a^{\dag}_{R}& = & \frac{a^{\dag}_{1}+i a^{\dag}_{2}}{\sqrt 2}\,\,\,,
\label{mlett:a45}\\
a^{\dag}_{L}& = & \frac{a^{\dag}_{1}-i a^{\dag}_{2}}{\sqrt 2}\,\,\,.
\label{mlett:b45}
\eea
\eml

\noindent
From (\ref{44}) one sees that, in a generic state, $Q$ counts the number
of left minus right polarized photons. It is easily checked that $Q$ commutes
with all the generators of the conformal group as should be the case
for an internal symmetry generator.

The last part of this Section is dedicated to present the functional
quantization of the Schwarz-Sen theory in the Coulomb gauge. Clearly, the
constraints (\ref{mlett:a14}) and (\ref{mlett:a18}) can be used to eliminate 
the phase-space
variables $\pi^a_0$, $A^{a,0}$ from the outset. On the other hand, the
constraints $\Omega^{a} \approx 0$, $\chi^a \approx 0$ have vanishing Poisson
brackets with those in the set $\{\Omega^a_{Ti}\approx 0\}$. This means
that the Faddeev-Popov determinant split as follows

\be
\label{450}
\det(\nabla^2)\,\,{\det}^{1/2}(\e^{ab}\e^{ijk}\pa_j)\,\,\,,
\ee

\noindent
which after taking into account the functional relationship

\be
\label{451}
{\det}^{1/2} (\epsilon_{ab} \epsilon_{ijk}\partial^j) =
{\det}(\epsilon_{ijk}\partial^j)\,\,\,,
\ee

\noindent
reduces to

\be
\label{452}
\det(\nabla^2)\,\,\det(\e^{ijk}\pa_j)\,\,\,.
\ee

\noindent
Clearly, the first factor in (\ref{450}) is the determinant of the matrix whose
elements are $[\Omega^a(\vx),\chi^b(\vy)]_P$, while the second is the
determinant of the matrix whose elements are given at (\ref{15}). 
Hence, for the model under analysis, the phase-space generating functional of 
Green functions ($\tilde W$) is given by
 
\bea
\label{453}
\tilde W\, &=&\, \int\, [{\cal D} A^{a,i}]\, [{\cal D}\pi^{a}_{i}] \,\delta
[\partial_i A^{a,i}]\,\delta [\partial^i\pi^{a}_{i}]\, \det(\nabla^2)
\nonumber\\
& \times & \delta [\pi^{a}_i +
\frac{1}{2} \epsilon^{ab} \epsilon_{ijk}\partial^j A^{b,k}]\,
\det (\epsilon^{ijk}\partial_j)\,{\rm e}^{i\tilde 
S_{eff}}\,\,\,,
\eea

\noindent
where

\be
\label{454}
\tilde S_{eff}\, = \,\int\, d^4x\,\left[ \pi_i^{a} {\dot A}^{ia} 
- \frac{1}{2}(\vec{\nabla} \times \vec A^{a})^2 \right]\,\,\,,
\ee

\noindent
is the corresponding effective action.

\section{\bf Local Duality Transformations for Maxwell Theory}

As is well known, for the free Maxwell field, in the Coulomb gauge, 
the phase-space Green functions generating functional is given 
by\footnote{This section is mainly based on Ref.\cite{Gi2}.}
 
\be
\label{46}
W\, =\, \int\, [{\cal D} A^i]\, [{\cal D}\pi_i] \, \det(\nabla^2)
\, \delta [\partial_i A^i]\, \delta [\partial^i\pi_i]
\, {\rm e}^{iS_{eff}}\,\,\,,
\ee

\noindent
where the effective action ($S_{eff}$) reads

\be
\label{47}
S_{eff}\,= \,\int d^4x\, \left[ \pi_i \dot A^i 
- \left(- \frac{1}{2} \pi^i\pi_i - \frac{1}{2} B^i B_i\right)\right]\,\,\,.
\ee

\noindent
Here, $B^i = -\e^{ijk}\pa_jA_k$ is the $i$-th component of the magnetic field,
while $\pi_i$ denotes the momentum canonically conjugate to $A^i$.
As shown in Ref.\cite{SD1}, up to surface terms, $S_{eff}$ remains invariant 
under the non-local duality transformations

\bml
\label{48}
\bea
&& A^i \rightarrow A^{\prime i}\,=\,A^i + \delta_D A^i;\,\,\,
\delta_D A^i\, = \,\theta \,\nabla^{-2}\, \epsilon^{ijk}\, \partial_j \pi_k 
\,\,\,,\label{mlett:a48}  \\
&& \pi_i \rightarrow \pi^{\prime}_i\,=\,\pi_i + \delta_D \pi_i;\,\,\,
\delta_D \pi_i\, =\, \theta \,\epsilon_{ijk} \,\partial^j A^k \,\,\,. 
\label{mlett:b48}
\eea
\eml

The point we would like to stress now is that these transformations can be 
made local by introducing the auxiliary fields $C^i$, and their corresponding 
canonical conjugate momenta $P_i$, defined as follows

\bml
\label{49}
\bea
&&\nabla^2 C^{i}\, = \, \epsilon^{ijk}\,\partial_j \pi_k\,\,\,,
\label{mlett:a49}\\ 
&&P_{i} =\epsilon_{ijk}\, \partial^j A^k\,\,\,,\label{mlett:b49}\\
&&\partial_i C^i\,= \, \partial^i P_i\,=\,0\,\,\,. \label{mlett:c49}
\eea
\eml

\noindent
In fact, from Eqs.(\ref{48}) and (\ref{49}) one verifies that

\bml
\label{50}
\bea
&& \delta_D A^i\, = \,\theta\, C^{i},\,\,\,\,\,\,\,
\delta_D C^i\, = \,-\,\theta\, A^{i}, \label{mlett:a50}\\
&& \delta_D \pi_i = \theta \, P_{i},\,\,\,\,\,\,
\delta_D P_i = \,-\,\theta \, \pi_{i}\,\,\,.\label{mlett:b50}
\eea 
\eml

Our next task consists in reformulating the Maxwell theory in terms of the 
fields $A^i$, $\pi_i$, $C^i$ and $P_i$. To this end, we first recall that

\bml
\label{51}
\bea
&&\left(\prod \delta [\,\,\,\,]\right)\,(\vec{\nabla} \times {\vec{C}})^2\,=\,
\left(\prod \delta [\,\,\,\,]\right)\,\pi_i \pi_i\,\,\,,
\label{mlett:a51}\\
&&\left(\prod \delta [\,\,\,\,]\right)\,\frac{1}{2} P_i {\dot C}^{i}\,=\,
\left(\prod \delta [\,\,\,\,]\right)\,\frac{1}{2} \pi_i {\dot A}^{i}\,\,\,,
\label{mlett:b51}
\eea
\eml

\noindent
where the following definition

\be
\label{52}
\prod \delta [\,\,\,\,]\,\equiv\,
\delta[\partial_i A^i]\, \delta[\partial^i\pi_i]\,
\delta[\partial_i C^i] \delta[\partial^i P_i]\,
\delta [C^i - \nabla^{-2} \epsilon^{ijk} \partial_j \pi_k]
\delta[P_i - \epsilon_{ijk}\partial^j A^k]\,\,\,,
\ee

\noindent
has been introduced. As consequence, 

\be
\label{53}
\left(\prod \delta [\,\,\,\,]\right)\,S_{eff}\,=\,
\left(\prod \delta [\,\,\,\,]\right)\,{\tilde {\tilde S}}_{eff}\,\,\,,
\ee

\noindent
where

\be
\label{54}
{\tilde {\tilde S}}_{eff}\,\equiv\,
\int d^4x\, \left\{ \frac{1}{2} \pi_i {\dot A}^{i}
+ \frac{1}{2} P_i {\dot C}^{i} 
- \frac{1}{2}\left[ \,(\vec{\nabla} \times {\vec{A}})^2\,+\,
 (\vec{\nabla} \times {\vec{C}})^2 \right]\right\}.
\ee

\noindent
Correspondingly, the Coulomb gauge generating functional $W$ of Maxwell 
theory can be cast as

\bea
\label{55}
W &=& \int [{\cal D} A^i]\, [{\cal D}\pi_i] \, 
\delta[\partial_i A^i]\, \delta[\partial^i\pi_i]\,\nonumber\\
& \times &[{\cal D}C^i]\,[{\cal D}P_i]\,
\delta[\partial_i C^i] \delta[\partial^i P_i]\,\det (\nabla^2)\nonumber\\ 
& \times & \delta [C^i - \nabla^{-2} \epsilon^{ijk} \partial_j \pi_k]
\delta[P_i - \epsilon_{ijk}\partial^j A^k]
{\phantom a} {\rm e}^{i{\tilde {\tilde S}}_{eff}}\,\,\,.
\eea

It is through this form of $W$ that we shall make contact with the
Schwarz-Sen model. Needless to say, $W$ in (\ref{55}) is, by construction, 
invariant under the set of local duality transformations (\ref{50}). We 
learnt in this Section that, by means of an appropriate enlargement of the
phase-space, one can incorporate duality as a local symmetry of sourceless 
electrodynamics.

\section{\bf Equivalence of the Maxwell and Schwarz-Sen Theories}

We have now at hand two duality symmetric theories. One is the Schwarz-Sen
theory, whose phase space is spanned by the variables $A^{1,i}$, $A^{2,i}$, 
$\pi^1_i$ and $\pi^2_i$. The other one is Maxwell theory, whose phase-space
variables are $A^i$, $C^i$, $\pi_i$ and $P_i$. We shall prove, in this
Section, that these theories are quantum mechanically equivalent\footnote{This
Section is mainly based on Ref.\cite{Gi2}}.

The initial step toward this proof consists in identifying those variables
whose behavior under infinitesimal duality transformations is the same. For the
coordinates, the task is easy. Indeed, under infinitesimal
duality transformations the Schwarz-Sen coordinates $A^{a,i}$ change as 
follows (see (\ref{103}))

\be
\label{56}
\delta_D A^{1,i}\,=\,\theta \,A^{2,i}, \,\,\,
\delta_D A^{2,i}\,=\,-\,\theta \,A^{1,i}\,\,\,.
\ee 

\noindent
Therefore, if one sets

\be
\label{57}
A^{1,i}\,=\,A^i\,\,\,, 
\ee

\noindent
one obtains, from (\ref{mlett:a50}) and (\ref{56}),

\be
\label{58}
A^{2,i}\,=\,C^i\,\,\,.
\ee

\noindent
The situation is slightly more involved for the momenta. By combining (\ref{mlett:b14}), 
(\ref{56}) and (\ref{57}) one finds

\be
\label{59}
\delta_D \pi^1_i\,=\,- \frac{1}{2} \e_{ijk} \pa^j \delta_D A^{2,k}\,=\,
\frac{\theta}{2} \e_{ijk} \pa^j A^{1,k}\,
=\,\frac{\theta}{2} \e_{ijk} \pa^j A^{k}\,\,\,.
\ee

\noindent
On the other hand, (4\ref{mlett:b14}) enables one to write

\be
\label{60}
\frac{\theta}{2} \e_{ijk} \pa^j A^{k}\,=\, \frac{1}{2} \delta_D \pi_i\,\,\,.
\ee

\noindent
Therefore,

\be
\label{61}
\pi^1_i\,=\,\frac{1}{2}\,\pi_i\,\,\,.
\ee

\noindent
Through a similar calculation, which uses (49.b) instead of (48.b), one arrives
to

\be
\label{62}
\pi_i^2\,=\,\frac{1}{2}\,P_i\,\,\,.
\ee 

We shall next take advantage of the identifications above to rewrite the 
generating functional of Maxwell theory in terms of the Schwarz-Sen 
variables. This change of integration variables in (\ref{55}) leads to 

\bea
\label{63}
W &=& \int [{\cal D} A^{a,i}]\, [{\cal D}\pi_i^a] \, 
\delta[\partial_i A^{a,i}]\, \delta[\partial^i\pi_i^a]\,\det (\nabla^2)
\nonumber\\ 
& \times &
\,\delta [A^{2,i} - 2\, \nabla^{-2} \epsilon^{ijk} \partial_j \pi^1_k]
\delta[2 \, \pi^2_i - \epsilon_{ijk}\partial^j A^{1,k}]\nonumber\\
& \times &
\,{\rm e}^{i\tilde S_{eff}}\,\,\,.
\eea 

\noindent
We emphasize that, when written in terms of the Schwarz-Sen variables, the 
Maxwell action (${\tilde {\tilde S}}_{eff}$) becomes the Schwarz-Sen action
(${\tilde S}_{eff}$). However, since the integration measure in (\ref{63}) 
does not appear to be that in (\ref{453}), a few more algebraic manipulations 
will be required to establish the equivalence between these theories. Let 
$L^{ik}$ be the differential operator 

\be
\label{64}
L^{ik}\,\equiv\,\e^{ijk} \pa_j\,\,\,.
\ee

\noindent
Then, 

\be
\label{65}
\delta[\pa_i A^{2,i}]\,A^{2,i}\,=\,\delta[\pa_i A^{2,i}]\,L^{ik} {\nabla}^{-2}
L_{km} A^{2,m}\,\,\,,
\ee
 
\noindent
which, in particular, implies that

\bea
\label{66}
&&\delta[\pa_i A^{2,i}]\,\delta [ A^{2,i}\,
-\,2\,\e^{ijk}\,{\nabla}^{-2}\,\pa_j
\pi_k^1 ]\nonumber\\
&=& \delta[\pa_i A^{2,i}]\,
{\det}^{-1}\left(-\e^{ijk} \pa_j \nabla^{-2} \right)\,
\delta[ \e_{klm} \pa^l A^{2,m}\,+\,2\,\pi^1_k]\,\,\,.
\eea

\noindent
On the other hand, from the eigenvalue equation

\be
\label{67}
\delta[\pa_j \phi^j]\,L^{ik} {\nabla}^{-2} L_{km}\,\phi^m\,=\,
-\,\delta[\pa_j \phi^j]\,\phi^i\,\,\,,
\ee

\noindent
one learns that

\be
\label{68}
\det \left( L^{ik} {\nabla}^{-2} L_{km} \right)\,=\,constant\,\,\,,
\ee

\noindent
or, equivalently,

\be
\label{69}
{\det}^{-1}\left(-\e^{ijk} \pa_j \nabla^{-2} \right)\,=\,
constant \times \det \left(\e_{klm} \pa^l\right)\,\,\,.
\ee

\noindent
By going back with (\ref{66}) and (\ref{69}) into (\ref{63}) one finds that

\bea
W & = &constant \times
\int [{\cal D} A^{a,i}]\, [{\cal D}\pi_i^a] \,  
\delta[\partial_i A^{a,i}]\, \delta[\pa^i \pi_i^a]\,
\det(\nabla^2) \nonumber\\
& \times & \delta[ \pi^a_i 
+ \frac{1}{2} \e^{ab} \epsilon_{ijk}\partial^j A^{b,k}]
\det(\e^{ijk} \pa_j)
\nonumber\\
& \times &
\,\exp \left\{\,i \int d^4x\,\left[ \pi^a_i A^{a,i}\,-\,\frac{1}{2}\,
\left(\vec{\nabla} \times \vec{A}^a\right)^2 \right] \right\}\nonumber\\
&=& constant \times {\tilde W}\,\,\,.
\eea

The purported proof of equivalence between the Maxwell and the Schwarz-Sen
theories is now complete. 

\section{\bf Conclusions}

We started in this work by canonically quantizing the Schwarz-Sen theory in the
Coulomb gauge. The resulting theory turned out to be physically sensible.
The Lagrangian density defining the theory is not covariant but, nevertheless,
we were able of constructing a set charges verifying the Poincar\'e algebra.

Later, we showed that the phase space associated with the Coulomb gauge
formulation of Maxwell theory can be conveniently enlarged in order to 
accommodate duality as a local symmetry. 

By analyzing the behavior under duality transformation, we were able of 
identifying the phase-space variables of the Maxwell theory with those of the
Schwarz-Sen theory. It was then possible to demonstrate that the corresponding
Green functions generating functionals were the same. 

\end{document}